\documentclass[preprint,superscriptaddress,nofootinbib,preprintnumbers,amsmath,amssymb]{revtex4-1}
\usepackage{graphicx,amsmath}
\usepackage{hyperref}
\usepackage{amssymb}
\usepackage[usenames,dvipsnames]{xcolor}
\usepackage[normalem]{ulem}
\usepackage{xfrac}

\newcommand{\be}{\begin{equation}}
\newcommand{\ee}{\end{equation}}
\newcommand{\beq}{\begin{equation}}
\newcommand{\eeq}{\end{equation}}
\newcommand{\bea}{\begin{eqnarray}}
\newcommand{\eea}{\end{eqnarray}}

\newcommand{\pb}{\mathrm{pb}}
\newcommand{\fb}{\mathrm{fb}}
\newcommand{\MeV}{\mathrm{MeV}}
\newcommand{\keV}{\mathrm{keV}}

\newcommand{\GeV}{\mathrm{GeV}}
\newcommand{\ev}{\mathrm{eV}}
\newcommand{\gev}{\mathrm{GeV}}

\newcommand{\eg}{\textit{e.g.}\ }
\newcommand{\ie}{\textit{i.e.}\ }
\newcommand{\etc}{\textit{etc}\ }
\newcommand{\sd}{s_\delta }
\newcommand{\sa}{s_\alpha}
\newcommand{\sbe}{s_\beta }

\newcommand{\mo}{\ensuremath{\mathcal{O}}}

\newcommand{\ignore}[1]{}

\makeatletter
\def\gsim{\mathrel{\lower2.5pt\vbox{\lineskip=0pt\baselineskip=0pt
           \hbox{$>$}\hbox{$\sim$}}}}
\def\lsim{\mathrel{\lower2.5pt\vbox{\lineskip=0pt\baselineskip=0pt
           \hbox{$<$}\hbox{$\sim$}}}}
\makeatother

\begin{document}
\setlength{\unitlength}{1mm}

\noindent \makebox[11.5cm][l]{\small \hspace*{-.2cm} }{\small Fermilab-Pub-17-468-T}  
\title{Light Signals from a Lighter Higgs}

\author{Patrick J. Fox}
\email{pjfox@fnal.gov}
\affiliation{Theoretical Physics Department, Fermilab, Batavia, IL 60510, USA}
\author{Neal Weiner}
 \email{neal.weiner@nyu.edu}
 \affiliation{Center for Cosmology and Particle Physics, Department of Physics, New York University, New York, NY 10003}
 \affiliation{Center for Computational Astrophysics, Flatiron Institute, 162 Fifth Ave, New York, NY 10010, USA}

\begin{abstract}
With the Higgs search program already quite mature, there is the exciting possibility of discovering a new particle with rates near that of the SM Higgs. We consider models with a signal in $\gamma \gamma$ below the SM Higgs mass, taking the recent $2.9\, \sigma$ (local) CMS excess at 95 GeV as a target. We discuss singlet models with additional vectorlike matter, but argue that a Type-I two Higgs doublet model provides a more economical scenario. In such a setup, going into regions of moderate-to-strong fermiophobia, the enhanced $\gamma \gamma$ branching ratio allows signals from $VH$+VBF production to yield $\sigma \times BR_{\gamma\gamma} $ comparable to total SM rates. Light $H$ production can be dominated via rare top decays $t \rightarrow b H^+ \rightarrow b W^{*} H$, which provides an alternate explanation of the excess. We consider this in the context of other Higgs anomalies, namely the LEP Higgs excess near the same mass, and excesses in $t\bar{t}h$ searches at Tevatron and LHC. We find that with $140\, \gev < m_{H^+} < 160\, \gev$, $\tan \beta \sim 5$ and a coupling to gauge bosons of $\sin^2 \delta \sim 0.1$, such a scenario can produce all the excesses simultanously, where $tth$ arise from contamination from the rare top decays, as previously proposed.  An implication of the Type-I scenario is that any $\gamma \gamma$ excess should be associated with additional elements that could reduce background, including $b$-jets, forward jets or signs of vector boson production.
\end{abstract}

\maketitle

\section{Introduction}
The search for the Higgs boson was a tremendous undertaking. Not just at the LHC, but in the decades and experiments that preceded it. Results from LEP and the Tevatron provided the basis on which the multi-channel searches at the LHC proceeded. With the Higgs now discovered, and the LHC awaiting more luminosity - but little more energy - it is worth turning some attention to understanding what sorts of particles might still lie hidden in these data.

The simplest reason to pursue this is straightforward - at the LHC, one expects a massive increase in luminosity, and thus sensitivity, to new states, even with couplings well below $\mathcal{O}(1)$. The second reason is that throughout the search for the Higgs, there have been a variety of tantalizing bumps and excesses, many of which have lingered as open questions on the myriad exclusion plots presented over the years. For many, it is impossible not to at least consider whether these bumps might tell a consistent story of some new physics beyond the standard model.

Amongst these bumps comes the most recent result from CMS~\cite{CMS:2017yta}, which shows a small excess near 95 GeV in the diphoton channel.  We begin our discussion, in Section~\ref{sec:signals}, describing ways to explain this excess and then show, in Section~\ref{sec:global} how some of these approaches may also explain historical excesses from LEP and the Tevatron, as well as excesses in other channels at the LHC.  We conclude in Section~\ref{sec:discussion}.

\section{Signals of Light from a Lighter Higgs}\label{sec:signals}

Recently, the CMS collaboration, searching for $h\rightarrow\gamma\gamma$ with 35.9 $\fb^{-1}$ at 13 TeV has reported a 2.9 $\sigma$ (local) excess at 95.3 GeV \cite{CMS:2017yta}. The overall rate is consistent with a production cross section $\sigma_{pp\rightarrow H} BR_{H\rightarrow \gamma\gamma} \simeq 0.1 \pb$, which is similar to the SM rate expected at that mass. This excess has already drawn attention \cite{Mariotti:2017vtv,Crivellin:2017upt}. Of course, we know from LEP that no SM-like Higgs boson exists at that mass \cite{Barate:2003sz}. Thus, using this as a concrete target, one can ask is what sorts of models can create a diphoton resonance with a cross section approaching that of the SM Higgs boson at the LHC. 

\subsection{Higgs signals from singlets}
A simple example is that of a singlet scalar, $\phi$, which has been extensively discussed (see e.g., \cite{Goldberger:1999un,Giudice:2000av,Csaki:2000zn,Goldberger:2008zz,DeRujula:2010ys,Low:2010jp,Davoudiasl:2010fb,Fox:2011qc}).  To allow for production from gluon fusion at the LHC the singlet must couple to extra vectorlike colored matter, $y \phi \bar{\Psi} \Psi$, necessitating the introduction of many new degrees of freedom.  If that matter is also electrically charged, then the decay to photons is automatic. Such particles that can appear at near-SM rates - but are easily distinguished from a SM Higgs - have been referred to previously as Higgs friends~\cite{Fox:2011qc}. In another context, in a higher mass regime around 700-800 GeV, this has been referred to as the ``everybody's model'' \cite{Strumia:2016wys}.

The production cross section for such a particle can be related to its gluon decay width, and these quantities for a SM Higgs of the same mass,
\begin{equation}
	\sigma_{\phi} = \sigma_h \times \frac{\Gamma_{\phi\rightarrow gg}}{\Gamma_{h \rightarrow gg}}
\end{equation}
where 
\begin{equation}
	\Gamma_{\phi\rightarrow gg} = \frac{y^2 \alpha_s^2 N_\Psi^2 m_\phi^3}{72 \pi^3 m_\Psi^2}.
\end{equation}
Where we assume $N_\Psi$ copies of a vectorlike Dirac color triplet fermion, with mass $m_\Psi$.

The overall cross section times branching ratio is then:
\begin{equation}
\sigma(gg\rightarrow\phi\rightarrow \gamma\gamma) = \sigma_h \times \frac{\Gamma_{\phi\rightarrow \gamma \gamma}}{\Gamma_{h \rightarrow gg}}\end{equation}
where
\begin{equation}
	\Gamma_{\phi \rightarrow \gamma\gamma} = \frac{y^2 \alpha^2 N_\Psi^2 Q_\Psi^4 m_\phi^3}{16\, \pi^3 m_\Psi^2} =y^2 N_\Psi^2 \left(\frac{Q_\Psi}{\sfrac{2}{3}} \right)^4
	\left(\frac{m_\phi}{95\, \gev}\right)^3\left(\frac{200\,\gev}{m_\Psi}\right)^2 \times 0.5\,\keV~
\label{eq:phiwidth}
\end{equation}
for $N_\Psi$ fermions with charged $Q_\Psi$.  A 95 GeV SM-like Higgs boson has a gluon fusion production cross section of 76.3 pb and a width into gluons of 0.15 MeV.  Thus, 
\begin{equation}
	\sigma_\phi BR_{\phi \rightarrow\gamma\gamma} \approx 0.5 \pb \times y^2 N_\Psi^2 \left(\frac{Q_\Psi}{\sfrac{2}{3}} \right)^4 \left(\frac{200\, \GeV}{m_\Psi}\right)^2.
\end{equation}
Thus, a signal at the size seen at CMS is still possible, but it requires new light colored particles.  Even a new colored fermion as light as 200 GeV could have escaped detection so far at the LHC, if it decays predominantly into three jets~\cite{Dobrescu:2016pda}.
However, while it appears one can evade LHC bounds on colored particles and still have a sizable signal, it is certainly not economical to add new states both to observe and explain the production.  Moreover, although we have not yet discussed them, such a model cannot hope to easily explain the other Higgs related anomalies present in the data.

An alternative approach to adding a singlet and new colored fermions is instead to mix the singlet, $s$, with the Higgs boson. In such a case, the light mass eigenstate's couplings to SM fields will be proportional to some mixing angle $\sin\delta$  (hereafter $\sd$). The dominant production of $s$ will be through gluon fusion, but will occur at a rate suppressed by $\sd^2$ . Furthermore, as the dominant branching ratio ($s\rightarrow b\bar{b}$) is also proportional to the fermion coupling, the rate to $\gamma \gamma$ is independent of this mixing,
\begin{equation}
	\sigma(pp\rightarrow s\rightarrow\gamma\gamma) \simeq \sd^2 \,\sigma_{pp \rightarrow h} \times \frac{\Gamma_{s\rightarrow \gamma \gamma}}{\sd^2 \, \Gamma_{h\rightarrow bb}}.
\end{equation}
Thus, the rate to produce $s$ in the diphoton channel is directly proportional to the $\gamma\gamma$ width of $s$.  Achieving a rate comparable to the SM Higgs then requires $s$ having a diphoton width comparable to the SM Higgs (\ie\ $\Gamma(\gamma\gamma)\approx 0.5\,\keV$), which is not possible through mixing alone. As we can see from (\ref{eq:phiwidth}), this is possible, if $s$ has $\mathcal{O}(1)$ couplings to additional light fermions which have $\mathcal{O}(1)$ electric charge. 

\subsection{Type I Two Higgs Doublet Models}\label{subsec:TypeI}

Perhaps the most economical model is the Type I two Higgs Doublet model (see discussion in \cite{Gunion:1989we}).  This model consists of two $SU(2)$ scalar doublets, $\Phi_{1,2}$ which have opposite charge under a discrete $\mathbb{Z}_2$ symmetry, we take both to have hypercharge $Y=1/2$.  All right-handed SM fermions are even under the $\mathbb{Z}_2$ which means that one doublet, $\Phi_1$, is fermiophobic.  Such a model provides some additional freedom in its couplings to gauge bosons and fermions, and containing already a charged scalar which can mediate new processes.  

We parametrize the two doublets as
\be
\Phi_1 = \begin{pmatrix}
- H^+ s_\beta +  G^+ c_\beta \\
\frac{1}{\sqrt2} \left(v c_\beta -  h s_\alpha +  H c_\alpha - i A^0  s_\beta + i G^0 c_\beta\right)
\end{pmatrix}
\, ,
\Phi_2 = \begin{pmatrix}
 H^+ c_\beta + G^+ s_\beta \\
\frac{1}{\sqrt2} \left(v s_\beta + h c_\alpha + H s_\alpha + i A^0 c_\beta  + i G^0 s_\beta \right)
\end{pmatrix}~.
\ee 
With $h$ corresponding to the Higgs observed at 125 GeV, and $H$ its CP-even partner.  The tree-level couplings of the Higgs mass eigenstates to fermions, relative to the coupling of a SM Higgs are
\be
c_f^h = \frac{c_\alpha}{s_\beta}=c_\delta -\frac{s_\delta}{t_\beta}\,, \quad c_f^H = \frac{s_\alpha}{s_\beta}=-\left(s_\delta +\frac{c_\delta}{t_\beta}\right)\,, \quad c_u^{A^0}=-c_{d,\ell}^{A^0}=\frac{1}{t_\beta}~,
\ee
Where we have introduced the angle $\delta = \beta-\alpha-\pi/2$ to parametrize the deviation of the Higgs couplings from SM values~\cite{Alves:2012ez}.  Similarly the couplings to gauge bosons are
\be
c_V^h = s_{\beta-\alpha} =c_\delta\,,\quad c_V^H = c_{\beta-\alpha} = -s_\delta~.
\ee

The cross sections and widths of the new Higgs boson vary differently depending on these angles.  
Normalizing to a $m_H = 95\, \gev$,  there are first those that scale with $(s_\alpha/s_\beta)^2$,
\begin{align}
\label{eq:saoversbeqns}
&\sigma(gg) = 76.3\, \pb\, \times \left(\frac{s_\alpha}{s_\beta}\right)^2,  &  
&\sigma(t\bar{t}H) = 1 \pb\, \times \left(\frac{s_\alpha}{s_\beta}\right)^2 , &
&\sigma(b\bar{b}H) =1  \pb\, \times  \left(\frac{s_\alpha}{s_\beta}\right)^2, \\
&\Gamma(b\bar{b}) = 1.9 \MeV\, \times \left(\frac{s_\alpha}{s_\beta}\right)^2 ,&
&\Gamma(\tau^+\tau^-) = 0.2 \MeV\, \times \left(\frac{s_\alpha}{s_\beta}\right)^2 , &
&\Gamma(gg) = 0.15 \MeV\, \times \left(\frac{s_\alpha}{s_\beta}\right)^2~,\nonumber
\end{align}
and also those that are proportional to $s_\delta^2$,
\begin{align}
&\sigma(VBF) =  5 \pb\, \times s_\delta^2,&
&\sigma(WH) = 3.4\,\pb\, \times s_\delta^2 ,&
&\sigma(ZH) = 2\,\pb\, \times s_\delta^2, \nonumber\\
&\Gamma(WW^*) =  0.01\, \MeV\, \times  s_\delta^2 \,, &
&\Gamma(ZZ^*) =  1.6\, \MeV\,\times 10^{-3}\, s_\delta^2~.
\label{eq:sdeqns}
\end{align}
Finally, the diphoton partial width is
\be
\Gamma(\gamma\gamma) = \left|1.31 s_\delta + 0.31\frac{s_\alpha}{s_\beta} \right|^2 \times 3.27 \times 10^{-3}\, \MeV~.
\label{eq:diphotonwidth}
\ee

The couplings of the 125 GeV Higgs boson are constrained to lie close to SM values \cite{Craig:2015jba} which means that $|s_\delta| \lsim 0.4$.
We will consider regions of mild fermiophobia, where the coupling of $H$ to fermions is suppressed. Thus, for the remainder of this paper we will consider the region $\sd<0$, the region $\sd>0$ is the region of (mild) fermiophilia. We define the ratio $f_{FP}^2=\sd^2 /(\sa^2/\sbe^2 )$ as the ``factor of fermiophobia'' and consider fermiophobic regions to be those where $f_{FP}^2\gg1$.  In these regions, the branching ratio to diphotons is enhanced
\begin{equation}
BR_{H\rightarrow \gamma\gamma} \simeq f_{FP}^2 (I_{WW}(m_H) + \frac{1}{f_{FP}} I_{tt}(m_H))^2,
\label{eq:BRgamgam}
\end{equation}
where $I_{WW}(m_H)$ and $I_{tt}(m_H)$ are mass dependent functions resulting from loop integrals.  For a 95 GeV Higgs boson, $I_{WW}(95\,\GeV)\approx 0.05$ and $I_{tt}(95\,\GeV)\approx 0.01$, and in the fermiophobic regime the branching ratio to diphoton is $BR_{H\rightarrow \gamma\gamma} \simeq  3 f_{FP}^2\times10^{-3}$. The approximate relationship above assumes the $b\bar{b}$ decay still dominates the total width. 

  \begin{figure}[t] 
   \centering
   \includegraphics[width=0.6\textwidth]{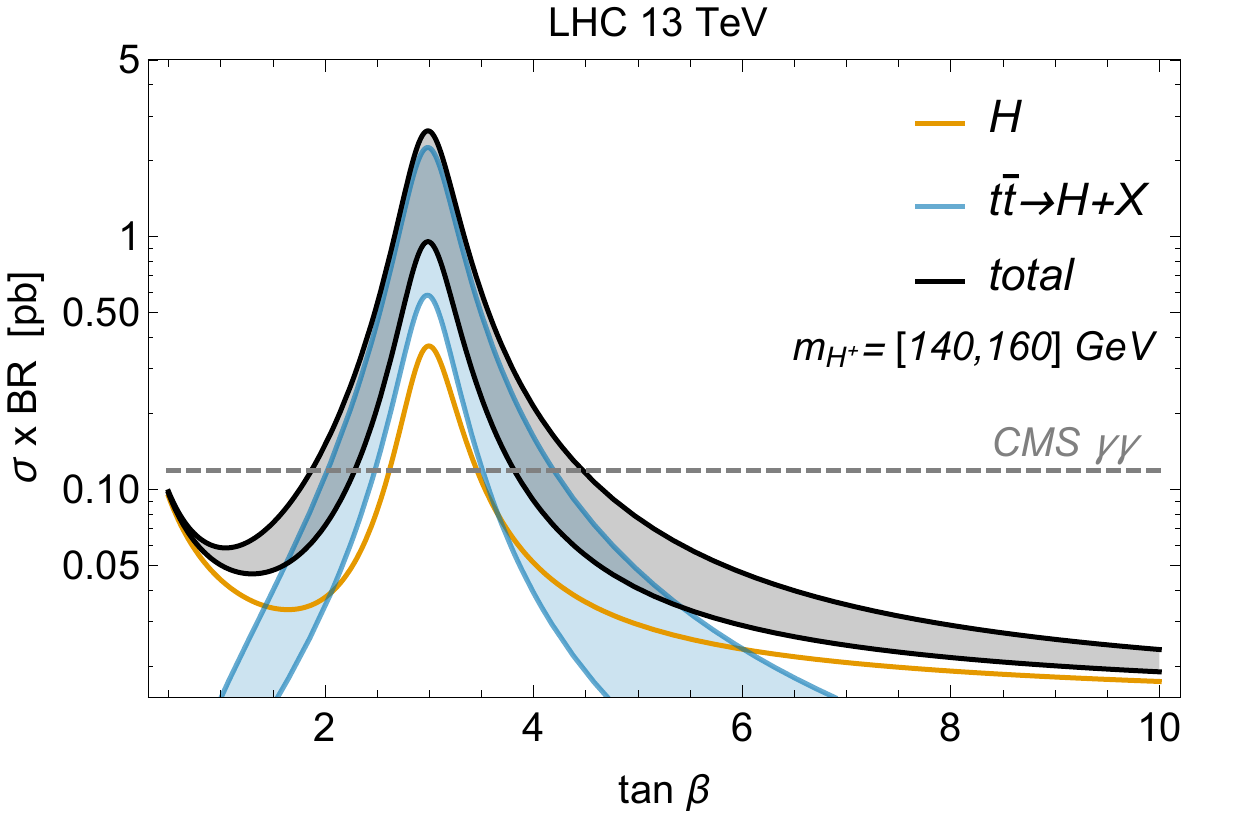} 
   \caption{Production rate, $\sigma_{pp \rightarrow H} \times {\rm BR_{H\rightarrow \gamma\gamma}}$, of a 95 GeV Higgs of at Type-1 2HDM at 13 TeV LHC assuming $\sd^2  = 0.1$. The lower (brown) curve is from all mixing induced processes (ggF, VBF, VH), the lower (blue) band is the contribution from rare cascade decays of tops with the charged Higgs mass in the range $[140,160]$ GeV, and the upper (black) band is the sum of both contributions. The dashed line shows the rate needed to explain the CMS excess at 95 GeV. }
   \label{fig:LHCrates}
\end{figure}

Using the results above, we see that, in the fermiophobic limit, the rate for $H\rightarrow \gamma\gamma$ through gluon fusion production scales as
\begin{equation}
	\sigma^{gg}_{pp \rightarrow H \rightarrow \gamma\gamma}  \approx 0.2\, s_\delta^2\,\pb~.
\label{eq:Hgg}
\end{equation}
With various bounds limiting $\sd^2 \lesssim 0.1$, this is a small fraction of the needed rate.  In contrast, VBF/VH production processes scale as
 \begin{equation}
	\sigma^{(VBF/VH)}_{pp \rightarrow H \rightarrow \gamma\gamma} = 0.03\, s_\delta^2\,f_{FP}^2 \,\pb~.
\label{eq:HVBF}
\end{equation}
For $f_{FP}^2\gsim 7$, the fermiophobic regime, gluon fusion will no longer be the dominant production channel for $H\rightarrow \gamma\gamma$. For strong fermiophobia $f_{fp}^2 \sim 20 \text{--}40$, total cross sections of $\mathcal{O}(0.1\,\pb)$ are possible, thus explaining the CMS excess. We show the directly produced signal in Fig \ref{fig:LHCrates}, and see that it can be large enough to explain the CMS excess for values of $\tan\beta$ around 3.
 
 
This discussion has so far focused on tree-level changes to the Higgs BR to photons. However, the light Higgs also couples to the charged Higgs. One expects a loop of charged Higgses and a resulting contribution to the width
\begin{equation}
\delta \Gamma_{H\rightarrow \gamma\gamma} = 
\frac{ \alpha^2  m_H^3}{2304\, \pi^3 m_{H^{+}}^2}\left|\frac{d  m_{H^+}}{d  H}\right|^2 \approx \left|\frac{d  m_{H^+}}{d  H}\right|^2 \left(\frac{m_H}{95\, \gev}\right)^3 \left(\frac{140\, \gev}{m_{H^+}}\right)^2 40\, \ev
 \end{equation}
In the fermiophobic limit with $\sd^2 =0.1$, the width from top and W boson loops gives $\Gamma_{H\rightarrow\gamma\gamma} \simeq 0.6\, \keV$.   Thus, for $\frac{d  m_{H^+}}{d  H} \lsim \mathcal{O}(1)$, the charged-Higgs loop are smaller than the SM contributions. The exact size of this contribution is, however, is very model dependent, and depends upon which operators split the charged Higgs from the neutral ones.  Typically, $\frac{d  m_{H^+}}{d  H}<1$ and the loop corrections from $H^+$ are not large.

 \begin{figure}[t] 
   \centering
   \includegraphics[width=0.45\textwidth]{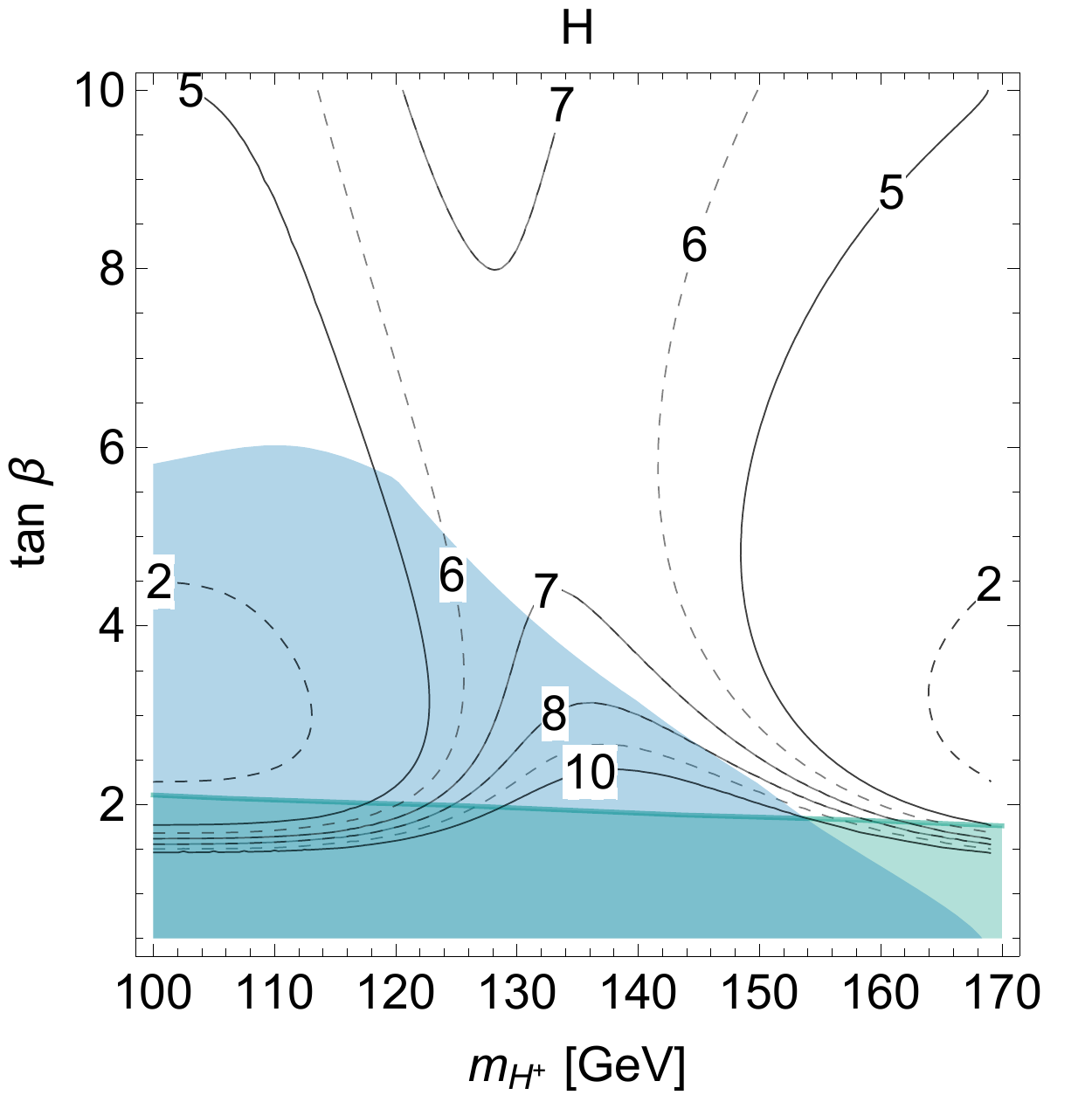}
   \includegraphics[width=0.45\textwidth]{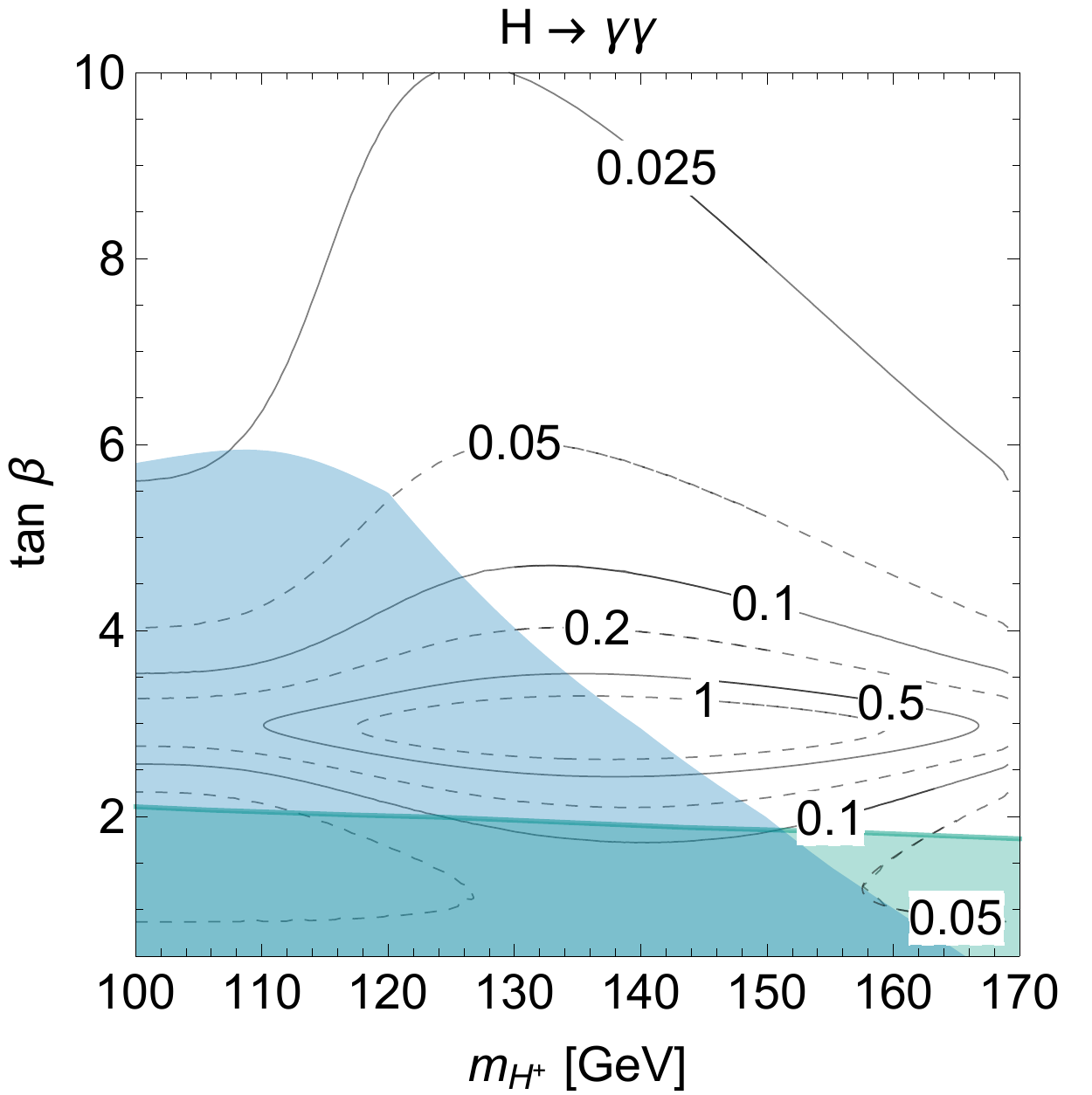}
   \caption{Contours of $H$ production cross section, $\sigma_H$, in pb (through direct $H$ production as well as production through top decay) at 13 TeV LHC, as a function of charged higgs mass, $m_{H^+}$ and $\tan\beta$ (left);  production cross section times branching ratio to $\gamma\gamma$, $\sigma_H BR_{H\rightarrow\gamma\gamma}$ (right).  In both figures the large (blue) region is ruled out by searches for rare top decays $t\rightarrow b\,(H^+\rightarrow \tau^+ \nu)$ and the smaller (green) region is ruled out by flavor observables ($\Delta M_s$ and $B_s^0\rightarrow \mu^+\mu^-$). We have taken the $A^0$ to be heavier than the top quark and $\sd^2=0.1$.}
  \label{fig:allowedmHptanbeta}
\end{figure}

There is another production possibility, again involving a light charged Higgs, that was recently emphasized by \cite{Alves:2017snd}. Namely, that that light scalar production can occur in cascades from a heavier charged Higgs \cite{Djouadi:1995gv,Akeroyd:1998dt,Coleppa:2014hxa,Coleppa:2014cca,Kling:2015uba,Arhrib:2016wpw}. In a type-I model, in the presence of a charged Higgs below the top mass, $m_{H^+}<m_t$, \cite{Alves:2017snd} showed that the production of the light Higgs via $t \rightarrow H^+ b \rightarrow H W^* b$, could be sizable, and consistent with existing constraints. For moderate $\tan\beta\lsim 6$, the branching ratio for this process can be $BR(t\rightarrow H W^* b)\sim \mathcal{O}(0.01)$, despite the decay of $H^+$ being three body. The top production cross section at 13 TeV is of 830 pb, which offers a $H$ production cross section of $\mathcal{O}(10\,\pb)$.  This can yield a CMS signal in the mildly fermiophobic regime (i.e., $f^2_{fp} \sim 5$), with $BR_{H\rightarrow\gamma\gamma} \sim 10^{-2}$. We show this combined signal in Fig \ref{fig:LHCrates}, and again there is a region at moderate $\tan\beta$ where the rate fits the CMS excess.  If this channel is available it will dominate production and one would expect additional signals in the LHC events.

There are several constraints on new light Higgs bosons that limit the available parameter space.  These constraints are weaker for a Type-I 2HDM than for Type-II.  Due to mass splittings among components of the Higgs doublets there are contributions to the precision electroweak observables $S$ and $T$, however these constraints are weak. There are indirect constraints from $B$-physics observables \eg\ $\Delta M_s$, $B_s^0\rightarrow \mu^+\mu^-$, $b\rightarrow s\gamma$, \etc~\cite{Enomoto:2015wbn}.  The strongest constraint over most of the parameter space we are interested in comes from searches for $t\rightarrow b\,(H^+\rightarrow \tau^+ \nu)$~\cite{CMS-PAS-HIG-14-020}.

\section{Global Perspective of Other Anomalies}
\label{sec:global}

With so many Higgs searches, it is perhaps not surprising that a number of anomalies have arisen.  Here we provide a brief discussion of a few of them and how one might attempt to explain them simultaneously.

\subsection*{LEP anomaly}
Using approximately 2.5 fb$^{-1}$ of data taken across a range of energies, $189\,\GeV <\sqrt{s}< 209\,\GeV$, the four LEP experiments searched for the process $e^+ e^-\rightarrow Z H$ where the Higgs boson decays into $b$ jets or tau leptons.  Combining all data~\cite{Barate:2003sz}, the experiments saw a broad excess ($>2\sigma$) above background expectations between 95 GeV and 100 GeV, with the largest deviation at $m_H= 99\,\GeV$.  LEP was most sensitive to $H\rightarrow b\bar{b}$ and this excess corresponds to a rate to $Z b\bar{b}$ of $\sim 0.1$ of the SM rate for a Higgs in the same mass range, \ie\ $\xi^2\equiv (g_{HZZ}/g^{SM}_{HZZ})^2\approx 0.1$.

\subsection*{LHC $\gamma\gamma$}
The CMS collaboration has carried out a search for diphoton resonances in the range $[80,110]$ GeV using 35.9 (19.7) fb$^{-1}$ of $\sqrt{s}=13\ (8)$ TeV data~\cite{CMS:2015ocq,CMS:2017yta}.  The combination of the two data sets has its largest discrepancy from SM background at $m_H=95.3$ GeV, corresponding to a local (global) significance of $2.8\sigma$ ($1.3\sigma$).  At this mass the 95\% confidence limit on the Higgs production cross section times branching ratio at $\sqrt{s}=13 (8)$ TeV is approximately 0.1 (0.05) pb.  At present ATLAS only has a search for diphoton resonances with $m_{\gamma\gamma}<110$ GeV for 20.3 fb$^{-1}$ of $\sqrt{s}=8$ TeV~\cite{Aad:2014ioa}, and no public analysis using $\sqrt{s}=13$ TeV data.  From this analysis the 95\% confidence limit on the Higgs production cross section times branching ratio at $m_{\gamma\gamma}\approx 95$ GeV is $\sigma BR \lsim 0.05$ pb.

\subsection*{LHC and Tevatron {$t\bar{t}H$}}

A variety of searches have been performed for a Higgs boson produced in association with a top quark pair. Notably, some of these have seen excesses~\cite{CMS-PAS-HIG-16-022,ATLAS-CONF-2016-058}, while others have not~\cite{CMS-PAS-HIG-16-038,ATLAS-CONF-2016-067,CMS-PAS-HIG-17-004,CMS-PAS-HIG-16-020}. Generally, more inclusive analyses (those employing cut-and-count approaches and sensitive to the specific value of $m_h=125\, \gev$) have seen greater excesses, while more exclusive analyses (those employing highly trained BDTs or neural nets or demanding $m_h=125\,\gev$) have not. We refer the reader to \cite{Alves:2017snd} for a thorough discussion.

A few important points are relevant, however: a search for $t\bar{t}H$ was performed by CDF~\cite{Collaboration:2012bk,Aaltonen:2013ipa} with $t\bar{t}H\rightarrow WW b \bar b b \bar b$ in the range $100\, \gev < m_h < 150\, \gev$. Because of the combinatorics of $b$ jets, the ability to discriminate a Higgs mass peak was poor. Nonetheless, the search shows a weak excess, reaching $\mo(2\sigma)$ near 100 GeV. 
At the LHC, ATLAS saw an excess in their multilepton analysis, which was cut and count, while CMS, employing a BDT did not. 

Critically, both experiments truncated their most focused $t\bar{t}H$ and VBF $h\rightarrow\gamma\gamma$ analyses at a point that a 95 GeV boson would have been missed (CMS has a lower bound of 100 GeV while ATLAS goes down to 105 GeV). The CMS low mass search \cite{CMS:2017yta} was sensitive to production $t\bar{t}H$ $Vh$, and VBF production mechanisms in addition to gluon fusion, but did not break out separate analyses for them, setting limits based on their expected relative rate and efficiencies in the SM.

\subsection{Explaining the Excesses with a Type-I 2HDM}

It is clear that one can explain any one of the excesses, for instance with a new singlet coupled to vectorlike fermions, but an intriguing question is whether one can explain most or even all of the excesses in a compact model.  As discussed in Section~\ref{subsec:TypeI} it is possible to explain the CMS $\gamma\gamma$ bump at 95 GeV in a Type-I 2HDM in the region of fermiophobia.  We shall argue that such a Type-I 2HDM provides a simple explanation for all excesses, while being consistent with null results.

The LEP results \cite{Barate:2003sz} are most simply understood as a type of scalar mixing with the Higgs boson at a level $\sd^2 \sim 0.1$. However, this could be a $SU(2)$ singlet or doublet scalar field.  Producing a $\gamma \gamma$ signal at the LHC comparable to the SM with such a small mixing is a challenge, however.  Absent new colored particles, one must boost the production cross section via mixing with the SM Higgs. Since such rates are necessarily below the SM, we must in turn resort to enhancing the $\gamma \gamma$ width of the new state.

\begin{figure}[t] 
   \centering
   \includegraphics[width=0.45\textwidth]{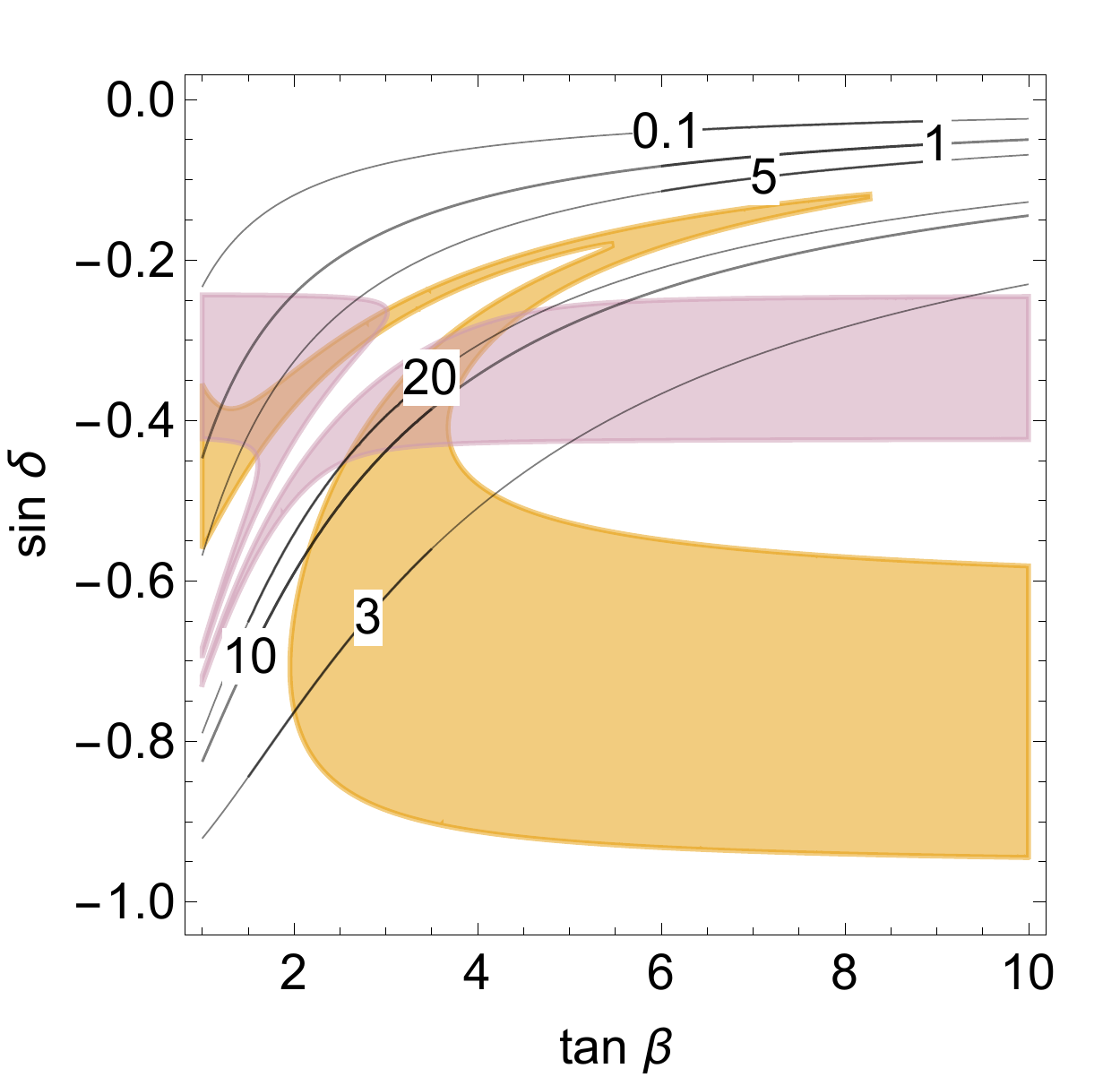}
   \includegraphics[width=0.45\textwidth]{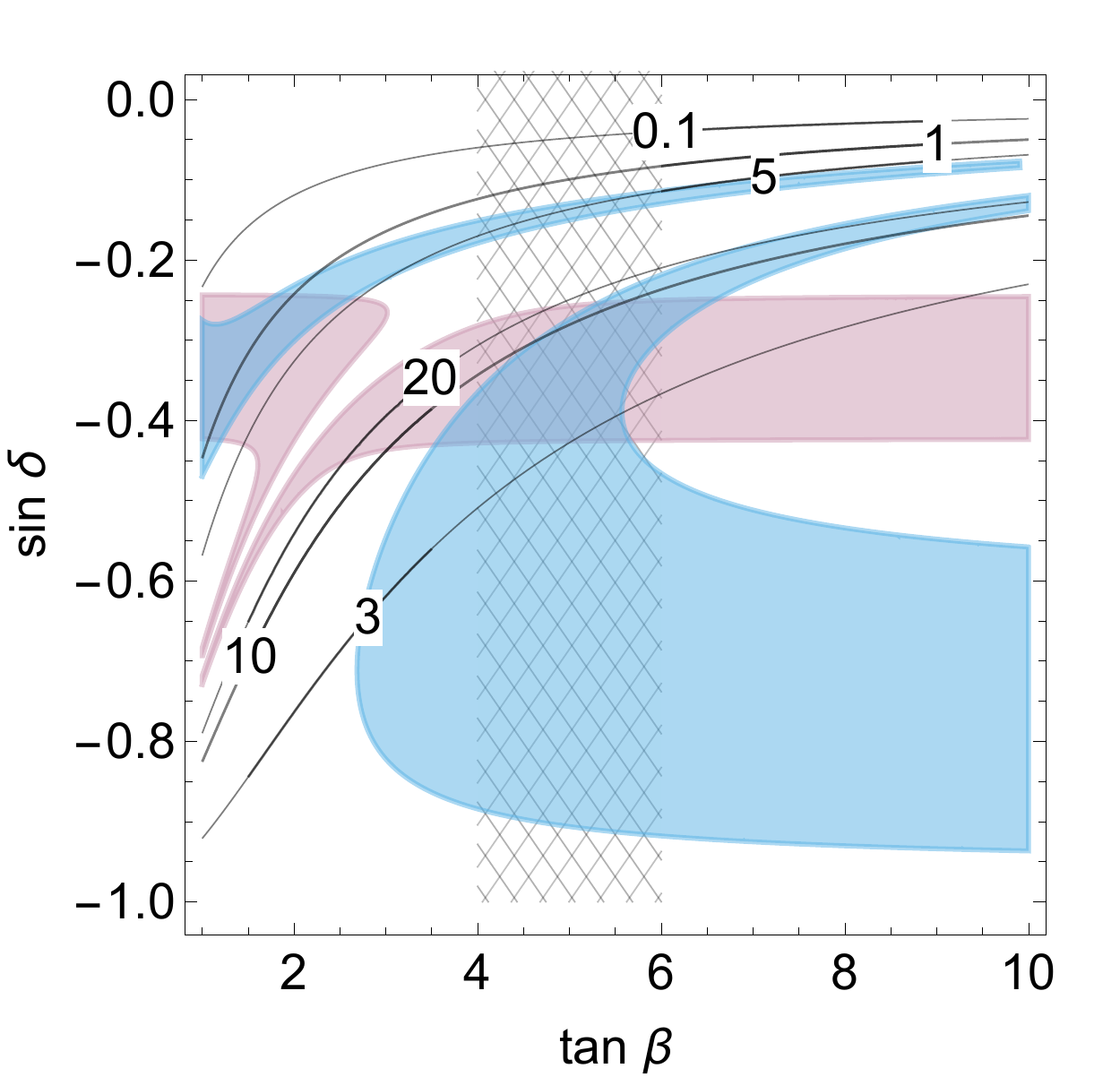}
   \caption{Contours of $f^2_{FP}$ as a function of $\sd$ and $\tan \beta$. Pink regions show areas consistent with the LEP excess ($0.05<\xi^2<0.15$) while the brown (left) or blue (right) region shows the area consistent with the CMS excess ($0.05\,\pb< \sigma BR_{H\rightarrow \gamma\gamma} <0.1\,\pb$). Left - H production arising only from ggF, VBF, VH processes. Right - H production including cascade decays from top quarks with $m_{H^+} = 140\, \gev$.  For the right plot, the approximate range (hashed) to explain the leptonic $t\bar t h$ excesses is $4 \lsim\tan \beta \lsim 6$ \cite{Alves:2017snd}. For $\sd > 0$, the Higgs is fermiophilic (i.e., $f_{fp}^2<1$) and the $ \gamma \gamma$ rates are suppressed.}
   \label{fig:globalrates}
\end{figure}

As shown earlier (\ref{eq:BRgamgam}), going to the fermiophobic regime, $f_{FP}\gg1$, increases $BR_{H\rightarrow \gamma\gamma}$.  With the requirement from LEP that $s_\delta^2\sim0.1$, we must go into the strongly fermiophobic regime, where $f^2_{fp} \approx 20-40$.   Then we find a signal at the LHC of $\sigma_{HV+HVBF} \times BR_{H\rightarrow\gamma\gamma}\sim 0.1\,\pb$, while the rate from gluon fusion is considerably smaller, (\ref{eq:Hgg}, \ref{eq:HVBF}).  That is, the CMS $\gamma\gamma$ excess can be explained not by $ggF$ but instead by the combination of VBF and associated production, which all lead to events with additional activity and other signals. We show the consistent region of parameter space in Fig.\ref{fig:globalrates} (left).

If the charged Higgs present in 2HDM's is lighter than the top mass there is an even more exciting possibility. This scenario, first discussed in \cite{Alves:2017snd}, has the dominant light Higgs production via $t \rightarrow H^+ b \rightarrow H W^* b$. 
It was argued in \cite{Alves:2017snd} that this process, involving a final state very similar to $t\bar{t}h$, would be a natural contaminant of those searches, and, indeed, could provide the explanation of the excesses seen. For a charged Higgs mass in the range $140\, \gev < m_{H^+} < 160\,\gev$, one needs $\tan\beta \approx 5$ to explain the $t\bar{t}h$ signals. Unfortunately, to explain the LEP excess in this $\tan \beta$ regime, one is naturally in the moderately fermiophobic regime, and is a non-trivial consistency check of this scenario. While it was noted by \cite{Alves:2017snd} that one could explain the LEP and $t\bar{t}h$ signals simultaneously, the near-inevitable boosted $\gamma \gamma$ signal was not recognized at the time. The global consistency of all three anomalies is shown in Fig.\ref{fig:globalrates} (right), for $m_{H^+}=140\,\GeV$. Note that increasing the charged Higgs mass shifts the required region for both CMS $\gamma\gamma$ and $t\bar{t}h$ to smaller $\tan\beta$.  This is compatible with the constraints on the $\tan\beta$ coming from rare top decays and indirect constraints from $B$ physics, see Figure~\ref{fig:allowedmHptanbeta}.  The same figure shows that to explain the anomalies there is a lower bound on the charged Higgs mass $m_{H^+}\gsim 130\,\GeV$ and an upper bound on the $H$ production cross section $\sigma_H\lsim 10\,\pb$.

If the top decay to $H^+$ is open then production of $H$ through top decay dominates over the sum of ggF, VBF, and VH meaning that there should be considerable additional activity in the excess $\gamma\gamma$ events \eg\ $b$ jets, leptons.  Furthermore, in the $t\bar{t}h$ searches there should also be a $\gamma\gamma$ resonance at 95 GeV.  Remarkably, the CMS and ATLAS searches for VBF and $t\bar{t}h$ with $h\rightarrow \gamma \gamma$ stopped short of going into this mass range.

Because the $BR(H\rightarrow\gamma \gamma$) is so much larger than in the SM, the expected rate is an order of magnitude - or more - beyond what is expected from the SM.  Assuming the efficiency to pass the analysis cuts for a 95 GeV Higgs is comparable to the SM one expects a considerable number of signal events just below the existing analysis range.  The resolution of the diphoton invariant mass is $\sim1.5$ GeV so a small fraction of the events centered around 95 GeV will leak into the analysis window, but this is too small to have been observed.  It almost defies belief, but the natural implication of this scenario is that there is an enormous signal lying just outside the currently searched mass window. While this seems unlikely, we cannot find any published paper or note that precludes this exciting possibility.

\section{Discussion}
\label{sec:discussion}
With the increasing sensitivity of Higgs searches, we confront the prospect of the discovery of new particles with Higgs-like properties. Simple models can provide signals into diphotons at rates comparable to the SM. Singlets can still provide high rates, but need additional light fields to provide production and/or widths to $\gamma \gamma$. In contrast, a doublet mixing with the SM Higgs in the form of a Type-I 2HDM provides an economical model that provides a boosted $\gamma \gamma$ signal in the fermiophobic regime of parameters. In the simplest case, the signal is generated not by $ggF$ but by $VBF+VH$ production, and thus would offer additional tags to improve separation of signal and background.

This last possibility is particularly exciting when viewed through the lens of a series of anomalies in Higgs searches. LEP ($ZH,H\rightarrow b \bar b$), CDF ($t \bar t H, H\rightarrow b \bar b$), ATLAS ($t \bar t H$, multilepton searches) and CMS ($H\rightarrow \gamma\gamma$) have all seen excesses consistent with a particle near $m_H = 95\, \gev$. The production follows the scenario proposed by \cite{Alves:2017snd}, where the light Higgs is produced in a cascade $t\rightarrow H^+ b \rightarrow b W^{+*} H$, which naturally contaminates the $t \bar t H$ searches. Interestingly, if one attempts to explain LEP along with $t \bar t H$ anomalies, one is inevitably forced into a region where the light Higgs is somewhat to very fermiophobic, and the $\gamma\gamma$ rate is enhanced. In such case, lowering the mass threshold for $t \bar t H$, $H\rightarrow \gamma \gamma$ searches, or looking for additional tags in conventional $H \rightarrow \gamma \gamma$ searches should yield dramatic signals well above SM rates. 

In summary, it is clear the prospect for discovery of new states in Higgs searches is significant. Moreover, if any of the anomalies above survice after further scrutiny and data, it may be that Higgs searches are not only the searches that completed the Standard Model, but may be the ones that find the first physics beyond it, as well.

\section*{Acknowledgments}
\label{sec:ack}

We thank Anna Maria Taki for helpful discussions.  The work of NW is supported by the Simons Foundation and by the NSF under grant PHY-1620727.  This work was supported by the DoE under contract number DE-SC0007859 and Fermilab, operated by Fermi Research Alliance, LLC under contract number DE-AC02-07CH11359 with the United States Department of Energy.  PJF would like to thank the CCPP and NYU for kind hospitality while this work was initiated. The Flatiron Institute is supported by the Simons Foundation.

\bibliography{lighthiggs}
\end{document}